\documentclass[prl,twocolumn,amsmath,amssymb,showpacs,superscriptaddress,groupedaddress]{revtex4}
\usepackage{graphicx}
\usepackage{dcolumn}
\usepackage{bm}
\usepackage[FIGTOPCAP]{subfigure}
\usepackage[usenames]{color}
\usepackage[all]{xy}
\usepackage{hyperref}
\usepackage{xcolor}
\hypersetup{
	colorlinks,
	linkcolor={red!90!black},
	citecolor={black!10!blue},
	urlcolor={blue!80!black}
}
\def \be{\begin{equation}}
\def \ee{\end{equation}}
\def \bea{\begin{eqnarray}}
\def \eea{\end{eqnarray}}

\begin{document}

\title{The random first-order transition theory of active glass in the high-activity regime}
\author{Rituparno Mandal}
\affiliation{Institute for Theoretical Physics, Georg-August-Universit\"{a}t G\"{o}ttingen, 37077 G\"{o}ttingen, Germany.}
\author{Saroj Kumar Nandi}
\affiliation{TIFR Centre for Interdisciplinary Sciences,
36/P, Gopanpally Village,
Serilingampally Mandal,
Hyderabad 500046, India.}
\author{Chandan Dasgupta}
\affiliation{Department of Physics, Indian Institute of
Science, Bangalore 560012, India.}
\author{Peter Sollich}
\affiliation{Institute for Theoretical Physics, Georg-August-Universit\"{a}t G\"{o}ttingen, 37077 G\"{o}ttingen, Germany.}
\affiliation{Department of Mathematics, King's College London, London WC2R 2LS, UK}
\author{Nir S. Gov}\affiliation{Department of Chemical and Biological Physics, Weizmann Institute of Science, Rehovot 7610001, Israel.}

\begin{abstract}
Dense active matter, in the fluid or amorphous-solid form, has generated intense interest as a model for the dynamics inside living cells and multicellular systems. An extension of the random first-order transition theory (RFOT) to include activity was developed, whereby the activity of the individual particles was added to the free energy of the system in the form of the potential energy of an active particle, trapped by a harmonic potential that describes the effective confinement by the surrounding medium. This active-RFOT model was shown to successfully account for the dependence of the structural relaxation time in the active glass, extracted from simulations, as a function of the activity parameters: the magnitude of the active force ($f_0$) and its persistence time ($\tau_p$). However, significant deviations were found in the limit of large activity (large $f_0$ and/or $\tau_p$). Here we extend the active-RFOT model to high activity using an activity-dependent harmonic confining potential, which we solve self-consistently. The extended model predicts qualitative changes in the high activity regime, which agree with the results of simulations in both three-dimensional and two-dimensional models of active glass.
\end{abstract}


\maketitle

\emph{Introduction.} Active glass is a condensed phase of matter that has internal sources of active (non-thermal) forces and extremely slow dynamics resembling in many ways the dynamics of passive glass-forming liquids. It has attracted a significant amount of interest as an abstract model for many biological systems \cite{angelini2011,nnetu2012,zhou2009,parry14,garcia2015,nishizawa2017,saroj2018} or synthetic soft active matter systems \cite{leomach19a,leomach19b,bartolo19}, and as a new challenge for non-equilibrium physics \cite{janssen2019active,berthier2019}. Realizations of active glass in numerical simulations are mostly in the form of a dense aggregate of interacting particles that are self-propelled; the self-propulsion appears in the form of random forces applied to each particle, characterized by a force amplitude $f_0$ and a persistence time $\tau_p$ \cite{ni2013,berthier2014,mandal2016}. A dimensionless quantity that is often used~\cite{Fily2014} to characterize the strength of activity is the ``active Peclet number'' $\mathrm{Pe} \equiv f_0\tau_p/(\gamma \sigma)$ where $\gamma$ is a friction coefficient and $\sigma$ is a microscopic length related to the particle size. Thus, the strength of activity can be increased by increasing $f_0$ or $\tau_p$.
Theories of equilibrium glasses have been extended for active systems and the resulting descriptions provide insights into many aspects of how activity affects the glassy properties~\cite{berthier2013non,szamel2016,liluashvili2017,feng2017,activemct,nandi2018random}. However, these theories are applicable in a regime where the activity is weak, i.e.\ $\mathrm{Pe}$ is small. Our aim in this work is to develop a theory for the regime of high activity.

We have recently presented an active random first order transition (active-RFOT) theory of an active glass \cite{nandi2018random}, where we proposed that the additional term in the free energy (or the ``effective temperature''), due to activity, is in the form of the potential energy of a single particle trapped in a harmonic potential \cite{ben2015modeling,wexler2020dynamics}. This ``effective medium'' model treats the particles that surround the test particle as a confining potential with spring constant $k$ and friction coefficient $\gamma$. The additional free energy term is given by \cite{nandi2018random}
\begin{equation}
\Delta F = \frac{H f_0^2 \tau_p}{1+k\tau_p/\gamma}
\label{dF}
\end{equation}
where $H=\tilde{H}/\gamma$, $\tilde{H}=T_K\kappa_a/\Delta C_p$, $T_K$ is the Kauzmann temperature, defined as the temperature where the configurational entropy of the passive system vanishes \cite{kauzmann1948nature}, $\Delta C_p$ is the jump in specific heat from the liquid to the crystalline state \cite{kauzmann1948nature}, and $\kappa_a$ is an active fragility parameter that quantifies the sensitivity of the configurational entropy to changes in the active force.
Using this modification we obtained an expression for the $\alpha$-relaxation 
time $\tau$\cite{nandi2018random},
\begin{equation}
\log{\left(\tau/\tau_0\right)}=\frac{E}{\left(T-T_K\right)+\Delta F}
\label{tau}
\end{equation}
where $\tau_0$ is a microscopic time scale and
$E$ represents the surface reconfiguration energy governing the relaxation dynamics of a region.
In the active-RFOT explored in \cite{nandi2018random} both $\tau_0$ and $E$ were assumed to be independent of the activity parameters $f_0, \tau_p$.

\begin{figure*}
\includegraphics[width=2.04\columnwidth]{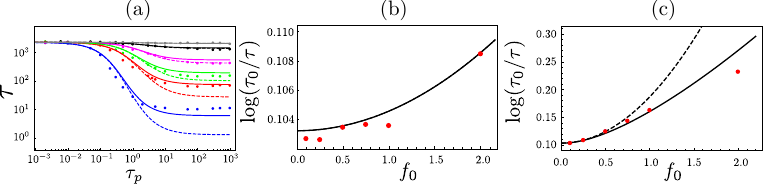} 
\caption{(a) Comparison between the measured $\alpha$-relaxation time $\tau$ from simulations of a three-dimensional active glass (at $T=0.45$) \cite{nandi2018random} (points) as function of $\tau_p$, and for different values of $f_0$: from top to bottom $f_0=0.1,0.25,0.5,0.75,1.0,2.0$. The dashed lines give the calculated relaxation time according to the active-RFOT  (Eqs.~\ref{dF}, \ref{tau}), using a constant $k=k_0$, while the solid lines use the mARFOT expression for the effective confinement $k=k_0+k_1\delta_{sc}$. (b,c) Plots of the simulated relaxation data of (a) as a function of $f_0$, plotted as the inverse of Eq.~\ref{tau} (red circles). Solid and dashed lines are the mARFOT and active-RFOT results, respectively. (b) $\tau_p=0.05$, so that $f_c\sim 70$ and we are in the quadratic regime of $f_0\ll f_c$. (c) $\tau_p=100$, so that $f_c\sim 3$, and indeed we see a clear deviation from the quadratic dependence, in agreement with the predicted $f_0^{4/3}$ power law. The values used in the fit are as in \cite{nandi2018random}: $\tau_0=0.135$, $T_K=0.29$, $E=1.55$, $H=0.042$, $k_0/\gamma=0.316$, with also $k_1=0.04$,$\gamma=1$.}
\label{fig1}
\end{figure*}

The theoretical expression (Eq.~\ref{tau}) was compared to simulation data for a three-dimensional active glass model \cite{nandi2018random}, and found to correctly predict the qualitative dependence of the relaxation time on the activity parameters (see Fig.~\ref{fig1}(a)). In the limit of large $\tau_p$, we find that the active term $\Delta F$ approaches a constant, but for large $f_0$ the predicted plateau values deviate significantly from the simulation results (see Fig.~\ref{fig1}(a)). Understanding the origin of this discrepancy at high activities is the central subject of this paper.

We assume that the basic RFOT phenomenology remains valid even at high activity and consider what modifications to the active term (Eq.~\ref{dF}) can account for this discrepancy. More specifically, at large activity, i.e.\ large active force $f_0$ and large persistence time $\tau_p$, the effective medium parameters $(k,\gamma)$ may become dependent on the activity. Since in the large $\tau_p$ limit the parameter $\gamma$ cancels out, we focus on the effective confinement parameter $k$. 
The surface reconfiguration energy, $E$, should also depend on activity, but fits of simulation data suggest that the modification of $E$ due to activity remains small even in this regime and we treat $E$ as a constant.

\emph{Activity-induced correction to the confinement.} Inside a dense active glass each particle is confined within a potential well formed by its neighbors. The active motion can lead to persistent squeezing of the particles against each other, leading to a stronger effective confinement due to the steep repulsive part of the inter-particle interaction potential. We now estimate the change in the effective confinement $k$ of a test particle in the glass due to the active fluctuations of the neighboring particles, using a density functional theory (DFT) formalism \cite{rytheory,hoell2019multi}.

For simplicity we do the calculation in one dimension, where the effective potential in DFT at $x$ has the form
\be
u(x)=\int_{-\infty}^{+\infty} d{x}^\prime v(x -x^\prime) [\rho(x^\prime) -\rho_l].
\label{potential1}
\ee
Here, $v(x)$ is the inter-particle potential (it is a function of $|x|$), $\rho(x)$ is the local number density field and $\rho_l$ is the density of the uniform liquid. In writing this equation, we have approximated the direct pair correlation function of the uniform liquid that appears in the Ramakrishnan-Yussouff form~\cite{rytheory} of the free-energy functional by $-v/(k_BT)$. The term involving $\rho_l$ is a constant that is ignored in the rest of the analysis. We consider two particles located at $x=\pm a$ and calculate $u(x)$ near $x=0$. The density near $x=\pm a$ is assumed to have a Gaussian form with variance $\delta$. The effective potential is given by
\begin{widetext}
\be
u(x)= \frac{1}{\sqrt{2\pi\delta}} \int_{-\infty}^{+\infty} dx^\prime v(x -x^\prime)\{\exp[-(x^\prime-a)^2/(2\delta)] + \exp[-(x^\prime+a)^2/(2\delta)]\}.
\label{potential2}
\ee
\end{widetext}
Clearly, $u(x) = u(-x)$, so that all odd derivatives of $u(x)$ at $x=0$ are zero. We need to calculate the second derivative of $u(x)$ at $x=0$: this gives the effective ``spring constant''. Consider the first term in Eq.~(\ref{potential2}):
\begin{eqnarray}
u_1(x) &=& \frac{1}{\sqrt{2\pi\delta}} \int_{-\infty}^{+\infty} dx^\prime v(x -x^\prime)\exp[-(x^\prime-a)^2/(2\delta)] \nonumber \\
&=&\frac{1}{\sqrt{2\pi\delta}} \int_{-\infty}^{+\infty} dy \,
v(x-y-a) \exp[-y^2/(2\delta)]
\label{u1}
\end{eqnarray}
where we have set $x^{\prime}-a=y$. The second derivative of $u_1(x)$ at $x=0$ is given by
\be
u_1^{\prime\prime}(x=0) = \frac{1}{\sqrt{2\pi\delta}} \int_{-\infty}^{+\infty} dy \,
v^{\prime\prime}(-y-a) \exp[-y^2/(2\delta)].
\ee
Now we make use of the fact that the exponential function is sharply peaked at $y=0$ because $\delta$ is small $(\delta \ll a^2)$, so that the main contribution to the integral comes from values of $y$ close to zero. This allows us to consider only the first non-vanishing term of the Taylor series expansion of $v^{\prime\prime}(-y-a) = v^{\prime\prime}(y+a)$ near $y=0$, giving
\begin{eqnarray}
u_1^{\prime\prime}(0) &=& \frac{1}{\sqrt{2\pi\delta}}
\int_{-\infty}^{+\infty} dy \, [v^{\prime\prime}(a) +\frac{1}{2} v^{\prime\prime\prime\prime}(a)y^2]\exp[-y^2/(2\delta)]. \nonumber \\
&=& v^{\prime\prime}(a) +\frac{1}{2} v^{\prime\prime\prime\prime}(a)\delta
\end{eqnarray}
Terms involving higher powers of $\delta$ can be obtained by retaining higher order terms in the Taylor series expansion of $v^{\prime\prime}(y+a)$. Combining this with the contribution of the second term in Eq.~(\ref{potential2}), we get
\be
u^{\prime\prime}(0) =  2 v^{\prime\prime}(a) + v^{\prime\prime\prime\prime}(a)\delta.
\ee
The effective confinement parameter therefore becomes
\begin{equation}
k=k_0+k_1\delta
\label{kx}
\end{equation}
where $k_0=2v^{\prime\prime}(a)$ and $k_1=v^{\prime\prime\prime\prime}(a)$.

\emph{Self-consistent calculation of the confinement.} The mean-square dispersion of the particles $\langle x^2 \rangle=\delta$ can now be calculated self-consistently. The potential energy of the confined active particle (Eq.~\ref{dF}) is related to its mean-square displacement: $\delta=2\Delta F/k$. Substituting Eq.~\ref{kx} into Eq.~\ref{dF} results in the following implicit equation for $\delta$
\begin{equation}\label{x2}
  \delta=\frac{2}{k_0+k_1\delta}\,\frac{Hf_0^2 \tau_p}{1+\left(k_0+k_1\delta\right)\tau_p/\gamma}
\end{equation}

%
%
The solutions of Eq.~\ref{x2} are the roots of a cubic polynomial in $\delta$
\begin{eqnarray}
a\delta^3+b\delta^2+c\delta&=&d\\ \label{cubic}
\begin{aligned}
a&=&k_1^2\tau_p/\gamma\\
b&=&k_1(1+2k_0\tau_p/\gamma)\\
c&=&k_0(1+k_0\tau_p/\gamma)\\
d&=&2Hf_0^2 \tau_p
\label{x2poly}
\end{aligned}
\end{eqnarray}


In the limit of large activity (force and persistence time), the discriminant of Eq.~\ref{cubic} is negative, and one obtains one real (and positive) solution for $\delta$, as well as two irrelevant complex solutions. For smaller activities, we have a single positive solution and two negative solutions that are discarded. 
Explicitly, we find that the self-consistent solution $\delta_{sc}$ initially increases linearly with $f_0^2$ for small $f_0$, but for large forces it now increases as $f_0^{2/3}$. The transition between these limits depends on the value of $\tau_p$:
\begin{equation}\label{delta}
\delta_{sc}\sim\left\{
\begin{array}{c l}	
     \frac{d}{c}=\frac{2Hf_0^2\tau_p}{k_0\left(1+k_0\tau_p/\gamma\right)} &,\ \quad f_0<f_c\\
     \left(\frac{d}{a}\right)^{1/3}=\left(\frac{2f_0^2H\gamma}{k_1^2}\right)^{1/3} &, \quad f_0>f_c
\end{array}\right.
\end{equation}
where the crossover force is given explicitly by $f_c=(k_0(\gamma+k_0\tau_p)/\tau_p)^{3/4}/(\sqrt{2Hk_1})$.

For $\tau_p\rightarrow\infty$ the critical force has the limiting value $f_c\rightarrow(k_0)^{3/2}/(\sqrt{2Hk_1})$. In this limit, the mean-square displacement of the particles can be written relatively compactly as
\be
\delta_\infty = \frac{k_0}{3k_1}
\frac{(\Lambda-1)^2}{\Lambda},
\quad
\Lambda = \left(1+d'+\sqrt{d'(2+d')}
\right)^{1/3}
\ee
where we have defined the dimensionless parameter $d'= 27 k_1 H\gamma f_0^2/k_0$.
In the limit of small $f_0$ we thus recover the quadratic dependence of $\delta_{\infty}$ on $f_0$, while at large $f_0$ the $f_0^{2/3}$ power-law of Eq.~\ref{delta} is obtained.

The self-consistent solution $\delta_{sc}$ can now be used in the modified confinement spring constant (Eq.~\ref{kx}), $k=k_0+k_1\delta_{sc}$, and in the active contribution to the free energy. With this modification we find that $\Delta F$ increases as $f_0^2$ for small $f_0$, but at large activity varies more slowly as $f_0^{4/3}$. We next compare this modified version of the active-RFOT (mARFOT) to simulation results.

\emph{Comparison to simulations.} Substituting the self-consistent solution we can compare to the simulation data of a three-dimensional active glassy system, shown in Fig.~\ref{fig1}~\cite{nandi2018random}. We find that the mARFOT greatly diminishes the discrepancy with the simulation data at large $\tau_p$ and $f_0$ (see Fig.~\ref{fig1}(a)). Note that these calculations use for the parameters $E,T_K,\tau_0$ values that are obtained from fits of $\tau$ for the passive system as a function of $T$, to the form of Eq.~\ref{tau} with $\Delta F=0$ \cite{mandal2016active,nandi2018random}. The values of the active parameters $k_0/\gamma,H$ are taken as in~\cite{nandi2018random}, determined by the fit at large $\tau_p$ and small $f_0$. This leaves only $k_1,\gamma$ as free fitting parameters.

\begin{figure*}
\includegraphics[width=2.04\columnwidth]{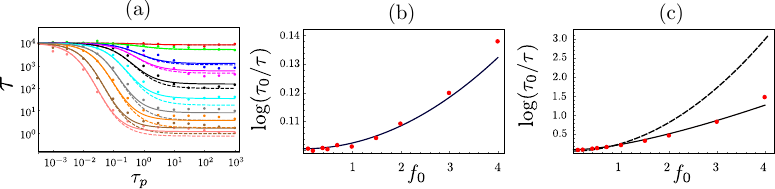} 
\caption{(color online) (a) Comparison between the measured $\alpha$-relaxation time $\tau$ from simulations of a two-dimensional active glass (at $T=0.4$, points) as a function of $\tau_p$, and for different values of $f_0$: from top to bottom $f_0=0.1,0.2,0.4,0.5,0.7,1,1.5,2,3,4$. The lines are for the active-RFOT (dashed) and mARFOT (solid). (b,c) Plots of the simulated relaxation data of (a) as a function of $f_0$, plotted as the inverse of Eq.~\ref{tau} (red circles). Solid and dashed lines are the mARFOT and active-RFOT results, respectively. (b) $\tau_p=0.01$, so that $f_c\sim 43$ and we are in the quadratic regime of $f_0\ll f_c$. (c) $\tau_p=1000$, so that $f_c\sim 1.4$, and indeed we see a clear deviation from the quadratic dependence, in agreement with the predicted $f_0^{4/3}$ power law. The other values used in the fit are: $\tau_0=0.5$, $T_K=0.2$, $E=1.99$, $H=0.4$, $k_0/\gamma=1.1$, with $k_1=0.8$,$\gamma=1$. }
\label{fig2}
\end{figure*}

Another way to expose the qualitative change at large $\tau_p$ is to plot $\log{\left(\tau_0/\tau\right)}$, i.e.\ the inverse of Eq.~\ref{tau}, as a function of $f_0$ for different values of $\tau_p$. This is shown  for two values of $\tau_p$  in Fig.~\ref{fig1}(b)-(c). It is clear that at low $\tau_p$ this function increases quadratically with $f_0$ (see Fig.~\ref{fig1}(b)), as predicted by the original active-RFOT expression (Eq.~\ref{dF}). For this value of $\tau_p=0.05$ we expect a crossover force $f_c\sim 70$ (Eq.~\ref{delta}), so we never enter the regime where the mARFOT is distinguishable. At a larger value of $\tau_p=100$ (see Fig.~\ref{fig1}(c)) we clearly find that the simulation data indicates an increase that is lower than quadratic, and is in good agreement with the predicted $f_0^{4/3}$ dependence of the mARFOT. For this value of $\tau_p$ the cross-over force is predicted to be $f_c\sim 3$, so the range of simulated forces does indeed enter the regime where mARFOT effects are significant.

In Fig.~\ref{fig2} we compare the model to new simulation data for a two-dimensional active glass (a finite temperature version of the model studied in~\cite{mandal2020extreme}). As in the three-dimensional case, we fit the $\alpha$-relaxation time of the passive system as a function of temperature to extract the parameters $T_K$, $\tau_0$ and $E$ (see SI, Fig.S1). The discrepancy between the simulation data and the active-RFOT is less strong compared to the data from the three-dimensional system (see Fig.~\ref{fig1}a), but it is clearly observable. The mARFOT is seen to resolve the major part of the discrepancy (see Fig.~\ref{fig2}(a)). It also captures the transition from quadratic (Fig.\ref{fig2}(b)) to non-quadratic dependence of the active term in the free energy on the force amplitude at high $\tau_p$ (Fig.\ref{fig2}(c)).

\emph{Discussion.} We provide here a self-consistent extension of the active-RFOT model that incorporates the renormalization of the effective confinement due to active fluctuations of the particles. This treatment gives a non-analytic modification of the power-law dependence of the active contribution to the free energy (``effective temperature'') on the active force magnitude. The predicted non-quadratic dependence, with power-law $f_0^{4/3}$, is in good agreement with our simulation data, both in three and two dimensions. This result may offer an explanation for similar puzzling observations in configurational entropy calculations~\cite{preisler2016}, where at high active forces a clear lower-than-quadratic dependence of the ``effective temperature'' on the active force was found.

Although our extension of the active-RFOT description improves the agreement with simulation results for high activity, there are regions in the parameter space $(f_0, \tau_p, T)$ where this theory is not expected to provide a very good description of the actual behavior. A recent study~\cite{mandal2020extreme} of an athermal ($T=0$) active system found a jamming transition as $f_0$ is reduced below a critical value $f_J$ in the $\tau_p \to \infty$ limit. For $f < f_J$ and large but finite values of $\tau_p$, the relaxation time is found to increase with increasing $\tau_p$. It is not clear whether this jamming transition would persist at moderate temperatures. However, our simulations for small values of $T$ and $f<f_J$ show a trend of increasing $\tau$ with $\tau_p$ for very large values ($\sim 1000$) of $\tau_p$. This behavior cannot be reproduced in the mARFOT description, which always predicts a decrease of $\tau$ with increasing $\tau_p$. Also, the active-RFOT and mARFOT descriptions cannot be used to describe the behavior for $T<T_K$. This is because the factor $T-T_K$  in Eq. \ref{tau}, arising from the temperature dependence of the configurational entropy for  $T > T_K$, should be replaced by zero for $T < T_K$. Eq.~\ref{tau} would then predict a temperature-independent relaxation time $\tau$ for $T<T_K$, but the (presumably weak) temperature dependence of parameters such as $\kappa_a, \gamma$ and $E$, which we have neglected in our treatment, would lead to values of $\tau$ that depend weakly on $T$. The temperature dependence of these parameters needs to be explored further. Finally, the effect of activity on the surface reconfiguration energy, ignored in our analysis, may become important for values of $f_0$ substantially higher than those considered here. Further examination of this effect would improve our present understanding of the properties of dense active matter in the high-activity regime.
Exploration of other effects of activity that are absent within our current framework, such as the one leading to motility induced phase separation~\cite{cates15}, would also be interesting.

\begin{acknowledgments}
N.S.G. is the incumbent of the Lee and William Abramowitz Professorial Chair of Biophysics, and acknowledges that this work is made possible through the historic generosity of the Perlman family. This project has received funding from the European Union’s Horizon 2020 research and innovation programme under Marie Sk\l odowska-Curie grant agreement No.\ 893128. SKN acknowledges support of the Department of Atomic Energy, Government of India, under Project Identification No. RTI 4007
\end{acknowledgments}

\bibliography{draft_ref}
\bibliographystyle{apsrev4-1}
\end{document}


\title{Supplementary Information: The random first-order transition theory of active glass in the high-activity regime}
\author{Rituparno Mandal}
\affiliation{Institute for Theoretical Physics, Georg-August-Universit\"{a}t G\"{o}ttingen, 37077 G\"{o}ttingen, Germany.}
\author{Saroj Kumar Nandi}
\affiliation{TIFR Centre for Interdisciplinary Sciences,
36/P, Gopanpally Village,
Serilingampally Mandal,
Hyderabad 500046, India.}
\author{Chandan Dasgupta}
\affiliation{Department of Physics, Indian Institute of
Science, Bangalore 560012, India.}
\author{Peter Sollich}
\affiliation{Institute for Theoretical Physics, Georg-August-Universit\"{a}t G\"{o}ttingen, 37077 G\"{o}ttingen, Germany.}
\affiliation{Department of Mathematics, King's College London, London WC2R 2LS, UK}
\author{Nir S. Gov}\affiliation{Department of Chemical and Biological Physics, Weizmann Institute of Science, Rehovot 7610001, Israel.}

\maketitle
\newpage

\subsection{Model for active glass in 2d}

For the 2d active glass simulations, we have used a previously studied active glassy system~\cite{mandal20, mandalprl20}. This model is essentially a 2d binary soft active sphere mixture at high density $\rho=1.2$. To implement the soft interaction between the particles (say between $i$ and $j$) we use a Lennard-Jones interaction
\begin{equation}
    V_{ij}(r)=4 \epsilon_{ij} \left[ \left(\frac{\sigma_{ij}}{r}    \right)^{12} -\left(\frac{\sigma_{ij}}{r} \right)^{6}\right].
    \label{pot}
\end{equation}
where inter particle distance is represented by $r$ and $\epsilon_{ij}$ and $\sigma_{ij}$ are the parameters representing the energy scale and the interaction radius, respectively. We have set all the energy scales in units of $\epsilon_{AA}$ and all the length scales in units of $\sigma_{AA}$ by choosing $\epsilon_{AB}=1.5 \epsilon_{AA}$, $\epsilon_{BB}=0.5 \epsilon_{AA}$, $\sigma_{AB}=0.8 \sigma_{AA}$ and $\sigma_{BB}=0.88 \sigma_{AA}$. We finally also set $\epsilon_{AA}=1$ and $\sigma_{AA}=1$. The passive limit of the model is known as Kob-Andersen glass~\cite{kob95,bruning08} and has been extensively used to study glassy dynamics in dense classical particle systems. To simulate the evolution we employ inertial dynamics with the equation of motion, 
\begin{equation}
m{\ddot{\mathbf{r}}}_i=-\gamma \dot{\bf{r}}_i + \mathbf{F}_{i} + f_0 \mathbf{n}_i +{\boldsymbol{\xi}}_i
\label{eqom}
\end{equation}
where $m$ is the mass of each particle and $\gamma$ is the friction coefficient. The position vector of the $i$-th particle is ${\bf{r}}_i$ where $i=1,\ldots,N$ and  $\mathbf{F}_{i}=-\nabla_i V_i$ is the total interaction force on particle $i$ derived from the LJ potential as $V_i=\sum_j V_{ij}$ (see Eq.~\ref{pot} for the definition). The interaction potential $V_{ij}$ has been truncated at  $R^c_{ij}= 2.5\sigma_{ij}$ and a quadratic smoothing function has been used so that both the energy and forces are continuous at $R^c_{ij}$. $f_0 \mathbf{n}_i$ is the active force acting on the $i$-th particle along the direction $\mathbf{n}_i$ associated with that particle. The unit vector $\mathbf{n}_i=\left(\cos{\theta_i}, \sin{\theta_i}\right)$ where $\theta_i$ is the angle representing the direction of the active force on the $i$-th particle. This angle (for each particle) is assumed to perform rotational Brownian motion following
\begin{equation}
   \dot{ \theta_i}=\sqrt{{2}/{\tau_p}}\, \zeta_i
    \label{eqom3}
\end{equation}
with rotational diffusion constant $\frac{1}{\tau_p}$. Here $\zeta_i$ is zero mean uncorrelated Gaussian white noise with correlator 
\begin{equation}
    \langle\zeta_i(t)\zeta_j(t^{\prime})\rangle=\delta_{ij}\delta(t-t^{\prime}).
\end{equation}
The thermal noise on the $i$-th particle is represented by ${\boldsymbol{\xi}}_i$ which has the property
 \begin{equation}
      \langle {\boldsymbol{\xi}}_i(t) \rangle=0 \,, \langle \xi^{\alpha}_i(t) \xi^{\beta}_j(t^{\prime})\rangle=2 \gamma k_B T \delta_{ij} \delta_{\alpha \beta} \delta(t-t^{\prime})
 \end{equation}
where $T$ is the temperature of the heat bath and $k_B$ is the Boltzmann constant. All the simulations were performed using a square periodic box with $N=1000$ particles. We used modified Langevin dynamics~\cite{beard00} for the MD simulation with $dt=0.005$. All the relevant quantities are averaged over time (after leaving out a transient time $t_T$ of $t_T=10^4$ which is $10$ times more than the largest $\tau_p$ studied) and also over $32$ independent simulations. To measure the relaxation timescale $\tau$ we first calculate the two-point overlap correlation function defined as $Q(t)$,
\begin{equation}
Q(t) =\frac{1}{N} \left\langle  \sum_{i} q(\mid{\bf r}_i(t)- {\bf r}_i(0)\mid)\right\rangle
\end{equation}
where $N$ is the number of particles in the simulation,
\begin{equation}
q(x)=
\left\{
        \begin{array}{ll}
                1  & \mbox{if } x \leq c\\
                0  & \mbox{otherwise}
        \end{array}
\right.
\end{equation}
and we have used $c=0.3$ (in units of $\sigma_{AA}$). The relaxation time $\tau$ is defined by $Q(\tau)=1/e$.
For the data presented in the main text, we have kept the temperature  at $T=0.4$ and extracted the time scale $\tau$ for different $f_0$ and $\tau_p$.

\subsection{Model for active glass in 3d}

For the three dimensional glass we have used data that has been shown in~\cite{nandi18} and has been generated from the model introduced in~\cite{mandal16}. The passive limit of the model (described in ~\cite{mandal16}) is a Kob-Andersen glass in 3d (a $80:20$ mixture of Lennard Jones particles with number density $\rho=1.2$). The parameters $\epsilon_{AA}$, $\epsilon_{AB}$, $\epsilon_{BB}$, $\sigma_{AA}$, $\sigma_{AB}$, $\sigma_{BB}$ are identical to the 2d Kob-Andersen model, described in the previous section.

Though the passive version of the 3d model is almost identical to the active glass model described in the previous section, in terms of implementation of activity this model is slightly different. For example all the particles in the previous model are active, whereas in this model only B particles were subject to a propulsion force, whilst keeping all the A particles passive. Self-propulsion forces are modelled inspired by an 8-state discrete Clock-model which has the form $\mathbf{f} = f_0 (k_x \hat{i} + k_y \hat{j} + k_z \hat{k})$, where $k_x,k_y,k_z$ are randomly assigned to have values $\pm 1$ (keeping the constraint that the net force is zero). The active particles are driven in the directions of $(k_x \hat{i} + k_y \hat{j} + k_z \hat{k})$ for a time scale (persistence time $\tau_p$) and then the directions of forcing are randomised by choosing a different set of $k_x, k_y , k_z$. Simulations are performed for $N = 1000$ particles in a 3d cubic box.

\subsection{Extraction of $E$, $\tau_0$ and $T_K$}

\begin{figure}
\centering
\includegraphics[width = \columnwidth]{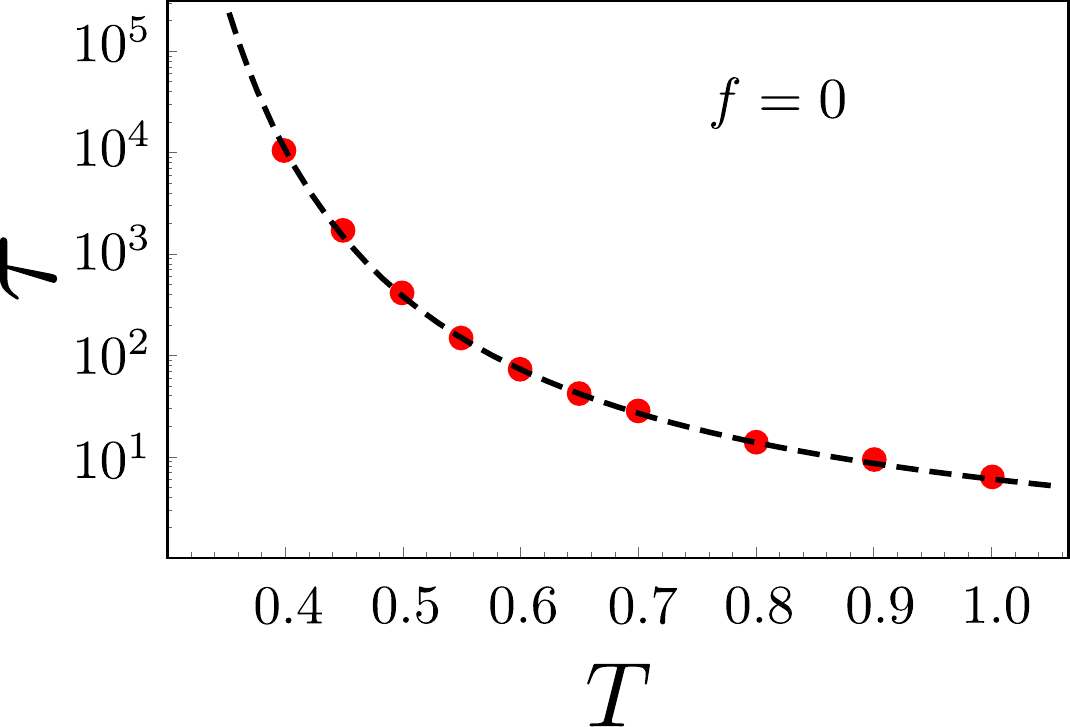}
\caption{Relaxation time $(\tau)$ has been plotted (red points) as a function of temperature $T$ for the passive system ($f=0$) in the two-dimensional simulations. The black dashed line shows a fit to the VFT form (Eq.\ref{tau_passive}).}
\label{fig:fitting}
\end{figure}

For the passive limit, \textit{i.e.} $\Delta F=0$ we have used a standard VFT (\tetxit{Vogel-Fulcher-Tammann}) form for fitting the relaxation time ($\tau$) versus temperature $T$ data. The VFT form is given by
\begin{equation}
    \tau=\tau_0 \exp{\left[ \frac{E}{T-T_K}\right]}
    \label{tau_passive}
\end{equation}
where $\tau_0$ is the microscopic time scale, $E$ represents the surface  reconfiguration  energy  governing  the  relaxation dynamics of a domain and $T_K$ is the Kauzmann temperature. This is the passive limit ($\Delta F=0$) of Eq.2. In Fig.~\ref{fig:fitting} we have shown a typical fit (dashed line) of the simulation data (red points) of the relaxation time which gives us the fit parameters $E$, $\tau_0$ and $T_K$.

\bibliography{si_ref}